\documentclass{article}
\usepackage{amssymb,amsmath,epsfig,epstopdf,color,graphicx,subfigure}
\usepackage{epstopdf,array}
\newcolumntype{L}{>{\centering\arraybackslash}m{3cm}}

\usepackage{arxiv}

\usepackage[utf8]{inputenc} % allow utf-8 input
\usepackage[T1]{fontenc}    % use 8-bit T1 fonts
\usepackage{hyperref}       % hyperlinks
\usepackage{url}            % simple URL typesetting
\usepackage{booktabs}       % professional-quality tables
\usepackage{amsfonts}       % blackboard math symbols
\usepackage{nicefrac}       % compact symbols for 1/2, etc.
\usepackage{microtype}      % microtypography
\usepackage[T1]{fontenc}
\usepackage[utf8]{inputenc}

\usepackage{amsmath,amssymb,amsfonts}
\usepackage{textcomp}
\usepackage{comment,color}
\usepackage{enumitem}
\usepackage{epsfig}
\usepackage{caption}
\usepackage{verbatim,bm}
\usepackage{epstopdf,array,mathtools,url,slashbox,subfigure}

\usepackage{booktabs}
\usepackage{soul,listings,float,amssymb,color,amsmath,epsfig,epstopdf,subfigure}
\usepackage[normalem]{ulem}
\usepackage{etoolbox}
\usepackage{lineno}
\usepackage{amsmath}
\usepackage{color}
\usepackage{algorithmicx,algorithm}
\usepackage[noend]{algpseudocode}
\usepackage{cancel}
\usepackage{lineno}

\newcolumntype{L}{>{\centering\arraybackslash}m{3cm}}
\DeclarePairedDelimiter\norm{\lVert}{\rVert}%
\makeatletter
\let\oldabs\abs
\def\abs{\@ifstar{\oldabs}{\oldabs*}}
\let\oldnorm\norm
\def\norm{\@ifstar{\oldnorm}{\oldnorm*}}
\makeatother
\title{Wavefield reconstruction inversion via physics-informed neural networks}

\author{
  Chao Song\\
 Physical Sciences and Engineering Division\\
 King Abdullah University of Science and Technology
    \And
 Tariq Alkhalifah\\
 Physical Sciences and Engineering Division\\
 King Abdullah University of Science and Technology
}

\begin{document}
\maketitle

\begin{abstract}

Wavefield reconstruction inversion (WRI) formulates a PDE-constrained optimization problem to reduce cycle skipping in full-waveform inversion (FWI). WRI often requires expensive matrix inversions to reconstruct frequency-domain wavefields. Physics-informed neural network (PINN) uses the underlying physical laws as loss functions to train the neural network (NN), and it has shown its effectiveness in solving the Helmholtz equation and generating Green's functions, specifically for the scattered wavefield. By including a data-constrained term in the loss function, the trained NN can reconstruct a wavefield that simultaneously fits the recorded data and satisfies the Helmholtz equation for a given initial velocity model. Using the predicted wavefields, we rely on a small-size NN to predict the velocity using the reconstructed wavefield. In this velocity prediction NN, spatial coordinates are used as input data to the network and the scattered Helmholtz equation is used to define the loss function. After we train this network, we are able to predict the velocity in the domain of interest. We develop this PINN-based WRI method and demonstrate its potential using a part of the Sigsbee2A model and a modified Marmousi model. The results show that the PINN-based WRI is able to invert for a reasonable velocity with very limited iterations and frequencies, which can be used in a subsequent FWI application.

\end{abstract}

% keywords can be removed

\section{Introduction}

Full-waveform inversion (FWI) is an effective technique to retrieve high-resolution subsurface properties of the Earth \cite{tarantola}. However, direct applications of FWI usually suffer from cycle skipping due to the sinusoidal nature of the wavefield and the complex scattering embedded in it \cite{vir}. The cycle-skipping issue can be reduced under the conditions that low-frequency components in the data are present or a good starting model is available. However, low-frequency data are often difficult to record and prone to be contaminated by noise. Researchers in our community have developed many methods to build a good initial model for FWI. Traveltime tomography extracts the first arrival information in the data to invert for a smooth velocity model in the shallow part \cite{clement}. Migration velocity analysis (MVA) can recover the kinematic features of the model by evaluating the quality of the image through, for example, measuring the flatness of angle-domain common-image gathers \cite{biondi,symes}. We can also extract specific features from the data to retrieve the background velocity updates, like seismic envelope inversion \cite{chi2014full,wu2014seismic,chen2020application}, phase-based waveform inversion \cite{bozdaug2011misfit,choi2013frequency}. Reflection-waveform inversion (RWI) is able to retrieve a smooth background model \cite{xu,zhou,wang2020enhancing}. RWI requires the calculation of a true-amplitude image at each iteration by a least-squares reverse time migration (LSRTM), which is computationally expensive. A reflection-based efficient wavefield inversion can improve the computational efficiency in calculating the perturbation model and mitigate cycle skipping in RWI, simultaneously \cite{song2020reflection}.

We can also seek a solution to cycle skipping from another perspective by formulating new objective functions instead of the conventional $l_{2}$ norm. In adaptive waveform inversion (AWI), \cite{warner2016adaptive,sun2020joint} proposed to transform the data using a convolutional operators and force Wiener filters to zero-lag delta function instead of the conventional data comparison. Wavefield reconstruction inversion (WRI) relaxes the wave equation accuracy to allow more data fitting and mitigate the nonlinearity of FWI \cite{van2013,van2015}. The key step of WRI and wavefield-reconstruction based methods is to reconstruct a frequency-domain wavefield, which simultaneously fits the data and satisfies the wave equation scaled by a weighting factor \cite{van2013,alkhalifah2019efficient,song2019tv}. This wavefield-reconstruction step needs to solve a so-called ``augmented wave equation'', which involves calculating a large matrix inverse. The computational cost and complexity increase dramatically for high frequencies and 3D large models, or complex physics, like anisotropy and elasticity. A fast method to reconstruct the frequency-domain wavefield is required to address these issues.

With the rapid developments in computing capability and the rampant availability of data, machine learning (ML) has gained wide attention in many fields. For example, the support vector machine (SVM) is a widely used technique for classification and regression \cite{vapnik2013nature}. SVM has shown its effectiveness in classifying seismic facies \cite{zhao2015comparison}, identifying passive seismic source type \cite{song2018source}, separating surface waves modes \cite{li2020separation}, suppressing artifacts in seismic images \cite{chen2020suppressing} and passive source images \cite{microsvm}. Neural network (NN) and its variation forms (e.g., deep/convolutional/recurrent NN) are becoming more powerful in pattern recognition, image processing and image segmentation for large-scale data. Deep NNs have shown effectiveness in picking first arrivals from raw seismic data \cite{zhu2019phasenet}, improve seismic image resolution by approximating a Hessian matrix inverse \cite{kaur2020improving} and FWI inverted velocity models by using well-log information \cite{zhang2020high,liyy,wang2020well}. Convolutional NN (CNN) has strong capabilities in extracting features from a large number of images, and it has been effectively applied to detect salt bodies \cite{shi2018automatic}, horizons, and faults from seismic images \cite{wu2019faultseg3d}, predict low-frequency components from high-frequency data \cite{ovcharenko2019deep}. Besides applications for analyzing the image features, CNN can also achieve seismic wave simulation in complex media \cite{moseley2020solving} and direct velocity inversions \cite{yang2019deep,zhang2020data,kazei2020velocity}. \cite{ren2020physics} implement FWI using a physics-based NN, but this method still suffers from the cycle-skipping issue in the conventional  FWI. \cite{8931232} develop SeisInvNets to build the mapping between seismic data and velocity models, and \cite{liu2021deep} further improve SeisInvNets by extracting more features from the data. However, SeisInvNets are only reliable for seismic data close to the training data. The above-mentioned ML-based velocity inversion methods are all based on time-domain seismic data, which require a large amount of training data. Besides applications in seismic exploration, ML-based methods have also been used in a variety of other geophysical problems, like fast ground penetrating radar (GPR) FWI \cite{giannakis2019machine}, electrical resistivity data inversion \cite{liu2020deep}, and earthquakes detection and prediction \cite{li2018machine,jiao2020artificial}.
  
ML has been successfully applied to solve PDEs by learning the physics disciplines from a large amount of input data. \cite{rudy2017data} proposed to apply a sparse regression method to solve PDEs using time series measurements in the spatial domain. NNs can be used as universal approximators to represent the general solutions of PDEs for complex datasets \cite{han2018solving,sirignano2018dgm,berg2019data}. A CNN is able to extract key features and simulate the dynamic evolution of underlying physics disciplines. \cite{tompson2017accelerating} propose a CNN-based framework to calculate fluid flow solutions corresponding to inviscid Euler Equations efficiently. CNN-based autoencoder is used as a surrogate modeling engine for a number of dynamical PDE systems \cite{geneva2020modeling}. The high fidelity and versatility of a CNN are also shown by \cite{thuerey2020deep} in solving for Reynolds-averaged Navier–Stokes (RANS) solutions. A new concept, referred to as Neural Operator, shows its effectiveness in learning the mapping between function spaces with limited training samples \cite{li2020neural}, and this method can be implemented in the Fourier space to accelerate the training process \cite{li2020fourier}. Most of the previous work based on SVM and NNs, referred to as supervised learning, tend to train a mapping system between input training data and their target outputs, which require a huge amount of data. Recently, \cite{raissi2019physics} proposed a physics-informed neural network (PINN) to solve partial differential equations (PDEs) instead of using a pure data-mapping loss objective. In PINN, we use the underlying physical laws in the loss function to reconstruct NN-based functional solutions to PDEs. Using the concept of automatic differentiation \cite{baydin2017automatic}, PINN can easily calculate the partial derivatives of NNs with respect to the input data, which often are spatial and temporal coordinates. In geophysical applications, PINNs have already shown effectiveness in solving the isotropic and anisotropic P-wave eikonal equation \cite{eikonal,waheed2020anisotropic}, Helmholtz equations for isotropic and anisotropic acoustic media \cite{helmholtz,song2020solving}. In these applications, the spatial coordinate values are used as input data, and the velocity and anisotropic parameters are considered as implicit parameters in the loss function. After training these NNs, we can evaluate the traveltime or the frequency-domain wavefield at any point in the domain of interest. \cite{waheed2021pinntomo} furhter develop a PINN-based tomography approach to invert for the velocity model. Besides geophysical applications, PINNs also show effectiveness in cardiac activation mapping \cite{sahli2020physics} and qualitative flow field characterization \cite{raissi2020hidden}. 

In generating the frequency-domain wavefields using PINN, \cite{helmholtz} proposed to solve the scattered form of the Helmholtz equation to avoid the point source singularity. An infinite isotropic homogeneous model is used as the background model to get analytical solutions of the background wavefield, which is computationally cheap and flexible of the model shapes. Since we can train NNs to satisfy the wave equation, we can also teach it to fit the data at the sensors' locations as well, yielding a WRI like implementation. In a seismic inversion, like FWI or WRI, a significant number of sources are required to provide enough illuminations to the subsurface. It is very expensive for PINN to generate the wavefield solution for every single source, as we need to train the same number as the source number independent PINNs. To improve the efficiency in generating the wavefields for multiple sources, \cite{alkhalifah2020machine} propose to add the horizontal source location as an extra dimension to the input data. As a result, we can use one NN to generate scattered wavefields for as many sources as required for different horizontal locations at a fixed depth, which is usually on the surface, like seismic surveys. After we reconstruct the wavefield reasonably well, we use another independent NN with a smaller size to predict the velocity. In this new network, we use the spatial coordinate values as input data, but the output is the velocity satisfying the scattered wave equation for the PINN-reconstructed wavefield at the input points. 

In this paper, we first review the concept of PINN. Then, we explain how we use two independent PINNs to achieve PINN-based WRI by estimating wavefields and inverting for the velocity, sequentially. Next, we show the results on a part of the Sigsbee2A model and a modified Marmousi model using PINN-based WRI. Finally, we discuss the potential advantage and also limitations of the proposed method.

\section{Theory}

\subsection{Physics-informed neural network}

Unlike supervised learning, which requires a large amount of data to evaluate the data-mapping based loss function, PINNs use principled physical laws as the loss function \cite{raissi2019physics}. By using the spatial and temporal coordinate values as input data, any order of partial differential equation can be easily evaluated with the help of automatic differentiation \cite{baydin2017automatic}. This is the frequency-domain acoustic wave equation with a constant density:
\begin{eqnarray}
\omega ^{2}m u(\mathbf{x},\omega)+\nabla^{2}u(\mathbf{x},\omega)=s(\mathbf{x_{s}}),
\label{eqn:eq1}
\end{eqnarray}
where $u(\mathbf{x},\omega)$ is the frequency-domain pressure wavefield; $\omega$ is the angular frequency; $m$ is the squared slowness. $s(\mathbf{x_{s}})$ is the source function, which is highly sparse, with $\mathbf{x_{s}}$ representing the spatial source coordinates. We can write Eq.\ref{eqn:eq1} in a compact form as: $L(m)u=s$. $L(m)=\omega ^{2}m+\nabla^{2}$ is the modeling operator. We define $f(\mathbf{x})=L(m)u-s$ to represent the physics-constrained loss function of NN to learn to approximate the wavefield $u(\mathbf{x},\omega)$. The mean squared error (MSE) of the physics-constrained loss is given by:
\begin{equation}
MSE_{f}=\frac{1}{N_{f}}\sum_{i=1}^{N_{f}}\left | f(\mathbf{x}_{f}^{i}) \right |_{2}^{2},
\label{eqn:eq2}
\end{equation}
where $ \left \{\mathbf{x}_{f}^{i}\right \}_{i}^{N_{f}}$ specify the selected input points from the domain of interest and $N_{f}$ is the selected point number. If there are recorded data on the boundaries or within the domain of interest to specify a unique solution for $u(\mathbf{x})$, we need to define a data-constrained MSE loss, which is given by:
\begin{equation}
MSE_{u}=\frac{1}{N_{u}}\sum_{i=1}^{N_{u}}\left | u(\mathbf{x}_{u}^{i})-u^{i} \right |_{2}^{2},
\label{eqn:eq3}
\end{equation}
where $ \left \{\mathbf{x}_{u}^{i}, u^{i}\right \}_{i}^{N_{u}}$ denote the recorded data on $u(\mathbf{x})$, and $N_{u}$ is the recorded data number. The total MSE loss is given by: $MSE=MSE_{f}+MSE_{u}$.

\subsection{The scattered wavefield}

We use the scattered wavefield, $\delta u=u-u_{0}$, as an alternative solution to get the wavefield. The Lippmann Schwinger form of the acoustic wave equation is shown as \cite{lippmann1950variational}:
\begin{eqnarray}
\omega ^{2}m \delta u+\nabla^{2}\delta u=-\omega ^{2}\delta mu_{0},
\label{eqn:eq4}
\end{eqnarray}
The scattered wavefield solution of Eq. \ref{eqn:eq4} is exact, not a Born approximation, and its source term on the right-hand side is associated with the background wavefield $u_{0}$ spanning the full space domain. The squared slowness perturbation is defined as $\delta m=m-m_{0}$, and $m_{0}$ represents the background squared slowness of a homogeneous model. The background wavefield $u_{0}$ satisfying the wave equation in Eq. \ref{eqn:eq1} with $m_{0}$, can be solved analytically. For 3D isotropic case, the analytical solution for constant velocity and a point source located at $\mathbf{x_{s}}$ is given by:
\begin{eqnarray}
u_{0}(\mathbf{x})=\frac{e^{i\omega \sqrt{m_{n0}}\left | \mathbf{x-x_{s}} \right |}}{\left | \mathbf{x-x_{s}} \right |},
\label{eqn:eq5}
\end{eqnarray}
where $\mathbf{x}=\left \{ x,y,z \right \}$ defines the spatial coordinates in the Euclidean space. For the 2D case, the analytical solution for the isotropic acoustic wave equation for a point source is expressed as:
\begin{eqnarray}
u_{0}(\mathbf{x})={iH_{0}^{(1)}(\omega \sqrt{m_{n0}}\left | \mathbf{x-x_{s}} \right |}),
\label{eqn:eq6}
\end{eqnarray}
where $H_{0}^{(1)}$ is the Hankel function of the first kind and order 0 \cite{engquist2018approximate}. Eqs. \ref{eqn:eq5} and \ref{eqn:eq6} are the analytical solutions to the Helmholtz equation corresponding to an infinite homogeneous isotropic medium. Given the true model, we can predict the scattered wavefield $\delta u$, in a similar way as the wavefield, using the physics-constrained $MSE_{f}$, which is defined as:
\begin{eqnarray}
MSE_{f}=\frac{1}{N_{f}}\sum_{i=1}^{N_{f}}\left | \omega ^{2}m_{f}^{i}\delta u_{f}^{i}+\nabla^{2}\delta u_{f}^{i}+\omega ^{2}\delta m_{f}^{i}u_{0f}^{i} \right |_{2}^{2}.
\label{eqn:eq7}
\end{eqnarray}

\subsection{The augmented wave equation}

In reality, the true velocity is often not available. In this case, we can reconstruct the wavefield using the augmented wave equation in WRI with a provided initial velocity  \cite{van2013}. The WRI objective function is given by:
\begin{eqnarray}
E(m,u_{is})= \sum_{is}^{N_{s}}\frac{1}{2}\left | d_{is} - u_{is}(\mathbf{x}=\mathbf{x_{r}}) \right |_{2}^{2}+ \\ \nonumber
\frac{\alpha}{2}\left |L(m)u_{is}-s_{is} \right |_{2}^{2},\\ \nonumber
\label{eqn:eq8}
\end{eqnarray}
where $is$ and $N_{s}$ represent the source index and total number, respectively. $d$ is the recorded data at the receivers' locations, which is denoted by $\mathbf{x_{r}}$. As a result, $u(\mathbf{x}=\mathbf{x_{r}})$ denotes the predicted data. In WRI, the wave equation is regularized by a weighting factor $\alpha$, which is set here to 0.00001. This value is chosen through trial and error tests and its role is discussed in \cite{alkhalifah2019efficient}. The WRI objective function in Eq. 8 fits the PINNs framework by using the data-fitting term as the data-constrained MSE and the regularized wave equation term as the physics-constrained MSE. As we are using the scattered wavefield instead of the whole wavefield, we need to fit the scattered data $\delta d$, which is obtained by subtracting the analytical background solution $u_{0}(\mathbf{x}=\mathbf{x_{r}})$ from the recorded data $d$. Thus, the total MSE loss to reconstruct the scattered wavefield using PINN is defined as:
\begin{eqnarray}
MSE=\sum_{is}^{N_{s}} ( \frac{1}{N_{r}}\sum_{i=1}^{N_{r}}\left | \delta d^{i} - \delta u(\mathbf{x_{r}}^{i}) \right |_{2}^{2}
+\\ \nonumber \frac{\alpha}{N}\sum_{i=1}^{N}\left | \omega ^{2}m_{1}^{i}\delta u^{i}+\nabla^{2}\delta u^{i}+\omega ^{2} (m_{1}^{i}-m_{0}^{i})u_{0}^{i} \right |_{2}^{2}), \\ \nonumber
\label{eqn:eq9}
\end{eqnarray}
where $N_{r}$ is the receiver number for the data constraint, and $N$ is the selected training point number for the physics constraint. In this network, the training input data are the spatial coordinate values $\mathbf{x}=\left \{x,z \right \}$ and horizontal source locations $x_{s}$. The target outputs are the real and imaginary parts of the reconstructed scattered wavefield $\delta u=\left \{\delta u_{r},\delta u_{i}\right \}$ corresponding to the initial squared slowness model $m_{1}$ for each source at regular grid points. The reconstructed scattered wavefields partially satisfy the scattered wave equation using the initial velocity model, while they also partially fit the recorded data at the receivers' locations simultaneously. Thus, the resulting reconstructed scattered wavefields will capture the information of the true velocity model thanks to the contribution of the data-fitting term. This information of the true velocity included in the reconstructed wavefields will be represented in the loss function of the network used to predict the velocity model and reflected in the inverted model. The velocity inversion step is explained in the next subsection. The background wavefield $u_{0}$ corresponding to the selected uniformly-distributed random points and source locations are calculated analytically. The initial squared slowness corresponding to the selected training points is calculated by linear interpolation from the regular grid they are defined at.  These parameters, including the background wavefield, initial and background model information, are implicit variables used to evaluate the loss function in Eq. 9, and their ordering must be consistent with the input data. We choose to optimize the loss function using an Adam optimizer with a stochastic gradient descent method \cite{kingma2014adam} and a follow-up L-BFGS optimization, a quasi-Newton approach, full-batch gradient-based optimization algorithm \cite{liu1989limited}. The learning rate for all the examples is 0.001. This training setup is used and suggested in the original PINN paper \cite{raissi2019physics}, and it is effective and efficient based on our experiments. A sufficient number of training epochs is needed to confirm the convergence, and we choose this number based on the observation of the loss curve through trial and error tests.

\subsection{The velocity inversion}

After reconstructing the scattered wavefield using PINN for an initial velocity, we build a new NN to predict the velocity. The input training data to this NN are the coordinate values of the regular grid points. In this NN, the target output is the squared slowness and it uses the PINN-reconstructed $\delta u$ as a known parameter. We use Eq. 7 as the loss function and add a total-variation (TV) regularization term to stabilize the training process, which is given by:
\begin{eqnarray}
MSE= \frac{1}{N_{p}}\sum_{i=1}^{N_{p}}\left | \omega ^{2}m^{i}\delta u^{i}+\nabla^{2}\delta u^{i}+\omega ^{2} (m^{i}-m_{0}^{i})u_{0}^{i} \right |_{2}^{2}\\ \nonumber
+\epsilon \sqrt{\left (  \frac{\partial m^{i}}{\partial x}\right )^{2}+\left (  \frac{\partial m^{i}}{\partial z}\right )^{2}}.\\ \nonumber
\label{eqn:eq10}
\end{eqnarray}
In this paper, we set $\epsilon$ equal to 0.1. The implicit variables $u_{0}$ and reconstructed $\delta u$ correspond to all the sources, so $N_{p}$ equals the number of regular grid points multiplied by the number of sources. For each frequency, we perform two tasks. The first task is to reconstruct the scattered wavefields, and the second task is to predict the velocity. We define these two tasks as one iteration. Finally, like frequency-domain FWI and WRI, we use multiple frequencies to refine the inverted velocity model. To sum up, the algorithm of PINN-based WRI is outlined in Algorithm 1.

\begin{algorithm}[t]
\caption{PINN-based WRI} %Algorithm caption
\hspace*{0.02in} {\bf Input:} %Input， \hspace*{0.02in}
Observed data $d$; True and background squared slowness models; Background wavefield $u_{0}$;  Iteration number $Niter$; Selected frequencies $[fmin:df:fmax]$; Weighting factor $\alpha$; \\
\hspace*{0.02in} {\bf Output:} 
 Inverted velocity model.
\begin{algorithmic}
\For{ifre=$[fmin:df:fmax]$;}
\For{iter=1:$Niter$;}
\State Reconstruct the scattered wavefields using PINN;
\State Predict the velocity with NN-reconstructed scattered wavefields using a velocity NN;
\EndFor
\State{\bf end for}
\EndFor
\State{\bf end for}
\end{algorithmic}
\end{algorithm}

\section{Numerical Tests}

In this section, we will show the results from implementing this approach on a part of the Sigsbee2A model and a modified Marmousi model. We refer to the meshing grid points representing the models as regular grid points. We generate the recorded data from a 9-point finite-difference (FD) wave equation solver \cite{jo1996optimal}. The source function we use for all the examples in this paper is a delta function in space. The networks are trained using a Quadro RTX 8000 GPU with 48 GB of memory.

\subsection{Sigsbee2A model}

\begin{figure}
\begin{center}
\includegraphics[width=1.0\textwidth]{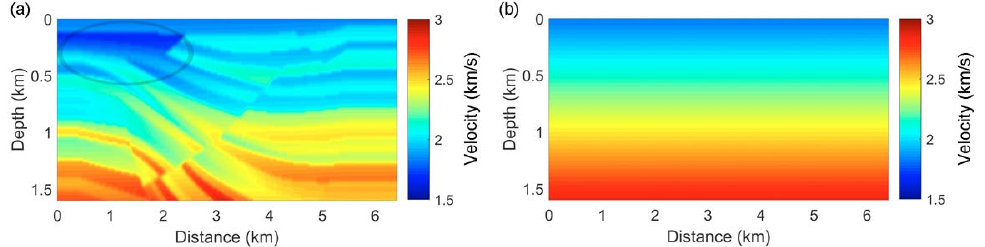}
\caption{The (a) true and (b) initial for the part of the Sigsbee2A velocity model. The horizontal ellipse region shows low-velocity zone in the shallow part.}
\label{fig:sigsbee_v_v0}
\end{center}   
\end{figure}

\begin{figure}
\begin{center}
\includegraphics[width=1.0\textwidth]{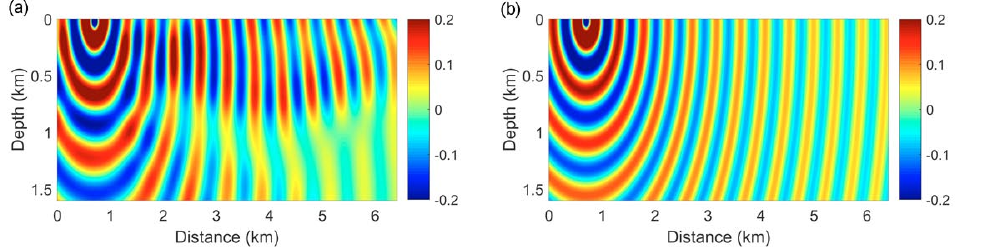} 
\caption{The real part of the 4 Hz (a) true wavefield calculated numerically corresponding to the true velocity in Fig.~\ref{fig:sigsbee_v_v0}a and (b) background wavefield calculated analytically corresponding to a homogeneous velocity of 1.83 km/s.}
\label{fig:sigsbee_u_u0}
\end{center}   
\end{figure}

\begin{figure}
\begin{center}
\includegraphics[width=0.5\textwidth]{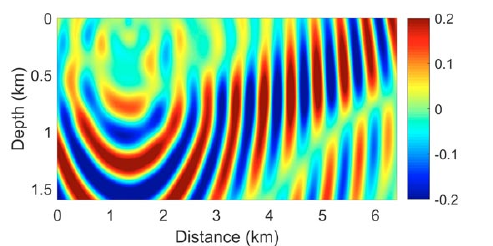} 
\caption{The real part of the 4 Hz scattered wavefield given by the difference between Figures~\ref{fig:sigsbee_u_u0}a and~\ref{fig:sigsbee_u_u0}b.}
\label{fig:sigsbee_du}
\end{center}   
\end{figure}

We first apply the proposed PINN-based WRI on a small part of the Sigsbee2A model. The true velocity model is shown in Fig.~\ref{fig:sigsbee_v_v0}a, and we observe a low-velocity zone in the shallow part, as shown in the horizontal ellipse. The initial velocity is linearly increasing with depth, as shown in Fig.~\ref{fig:sigsbee_v_v0}b. The size of the model is $401 \times 101$, and the vertical and horizontal spatial sampling interval is 16 m. We use 10 sources uniformly distributed on the surface, and the data are recorded at a depth of 16 m using 401 receivers. We implement the proposed PINN-based WRI just using one single frequency of 4 Hz to reconstruct the scattered wavefield and invert for the velocity. For the second source on the left, we show the real part of the true wavefield calculated numerically corresponding to the true velocity in Fig.~\ref{fig:sigsbee_u_u0}a and the background wavefield calculated analytically corresponding to a homogeneous velocity model of 1.83 km/s in Fig.~\ref{fig:sigsbee_u_u0}b. We calculate the wavefield difference between Figs.~\ref{fig:sigsbee_u_u0}a and~\ref{fig:sigsbee_u_u0}b to get the true scattered wavefield, as shown in Fig.~\ref{fig:sigsbee_du}.

\begin{figure}
\begin{center}
\includegraphics[width=0.5\textwidth]{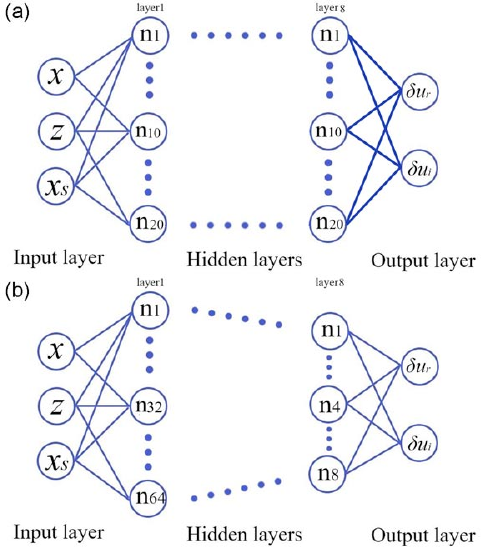} 
\caption{The (a) uniform and (b) hierarchically decreasing NN architectures used to reconstruct the scattered wavefield.}
\label{fig:DNN_sx_l20_l64}
\end{center}   
\end{figure}

\begin{figure}
\begin{center}
\includegraphics[width=1.0\textwidth]{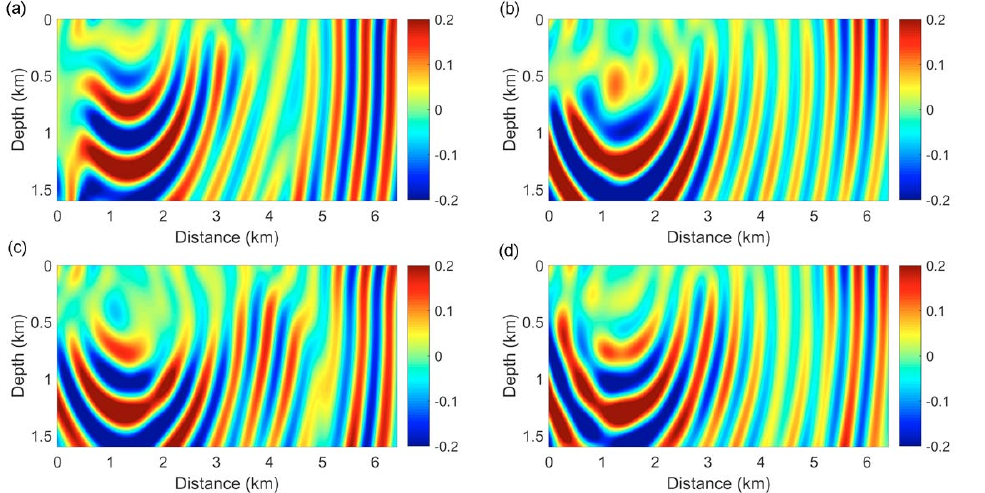} 
\caption{The real part of the 4 Hz PINN-reconstructed scattered wavefield from 8-layer uniform NNs with (a) 20, (b) 40, (c) 60 in each hidden layer, and (d) a hierarchically decreasing NN.}
\label{fig:sigsbee_du_ml}
\end{center}   
\end{figure}

\begin{figure}
\begin{center}
\includegraphics[width=1.0\textwidth]{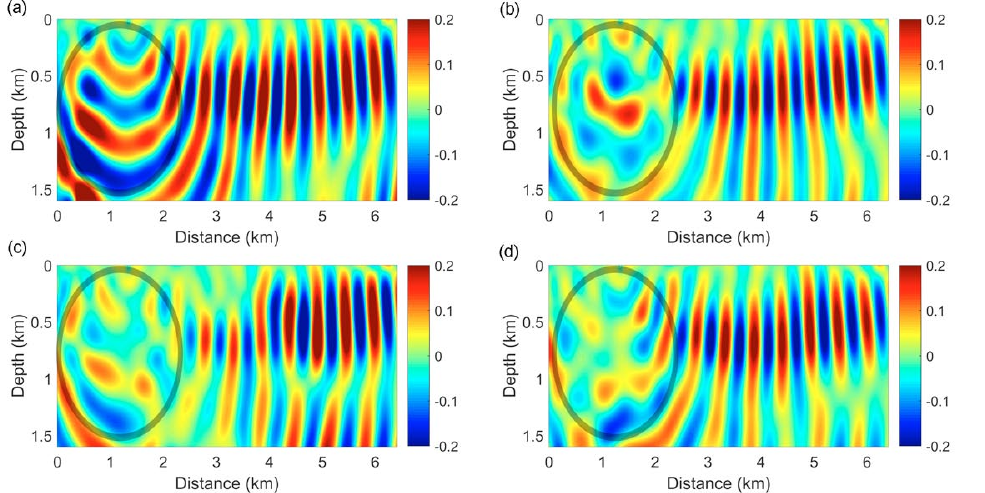} 
\caption{The real part of scattered wavefield difference between the true scattered wavefield in Fig.~\ref{fig:sigsbee_du} with the PINN-reconstructed scattered wavefield (a) in Fig.~\ref{fig:sigsbee_du_ml}a, (b) in Fig.~\ref{fig:sigsbee_du_ml}b, (c) in Fig.~\ref{fig:sigsbee_du_ml}d, and (d) in Fig.~\ref{fig:sigsbee_du_ml}d. The vertical ellipse regions show smaller scattered wavefield difference than the other regions around the source location.}
\label{fig:sigsbee_du_dif}
\end{center}   
\end{figure}

In this example, we generate 100000 random sets of $\left \{x_{i},z_{i}, x_{si}\right \}$ as the training data to feed to an 8-layer fully connected NN to reconstruct the scattered wavefield. We start with a small NN, which has 20 neurons in each hidden layer. The NN architecture is shown in Fig.~\ref{fig:DNN_sx_l20_l64}a. We refer to this NN architecture setup as a uniform network. We train this network for 20000 epochs of the Adam optimizer and 50000 iterations of LBFGS. The real part of the PINN-reconstructed scattered wavefield for the second source is shown in Fig.~\ref{fig:sigsbee_du_ml}a, and we observe that the left edge of the scattered wavefield is poorly reconstructed. To better verify the accuracy of the PINN-reconstructed scattered wavefield, we calculate the difference between the true and PINN-reconstructed scattered wavefields and show it in Fig.~\ref{fig:sigsbee_du_dif}a. Clearly, the scattered wavefield difference is large in the whole domain, even the areas near the source location in the vertical ellipse. These large differences indicate a poor recovery of the wavefield reconstruction using PINN with a small number of neurons. The runtime per epoch using the Adam optimizer is 0.113 s, and the runtime per iteration of LBFGS is slightly smaller than that of the Adam optimizer.

Then, we try to increase the size of the NN architecture by using more neurons in each hidden layer. Using 40 or 60 neurons in each hidden layer, the PINN-reconstructed scattered wavefields are shown in Figs.~\ref{fig:sigsbee_du_ml}b and~\ref{fig:sigsbee_du_ml}c, respectively. We see that the general shapes of the PINN-reconstructed scattered wavefields from larger NNs fit the true scattered wavefields well. The scattered wavefield difference corresponding to the network with 40 neurons in each hidden layer is much smaller than the case for 20 neurons, especially in the vertical ellipse, as shown in Fig.~\ref{fig:sigsbee_du_dif}b. Fig.~\ref{fig:sigsbee_du_dif}c shows further difference reduction using the 60-neuron uniform network. However, the more neurons used in each hidden layer, the more runtime required for training the network. The runtime per epoch using the Adam optimizer using 8-layer NNs with 40 and 60 neurons are 0.151 s and 0.213, respectively. To improve the computational efficiency, we use more neurons in the shallow layer, and gradually decrease the neuron number in the deep layers. We refer to this NN architecture setup as a hierarchically decreasing network. In this case, the neuron number in each hidden layer is $\left \{64, 64, 32, 32, 16, 16, 8, 8 \right \}$ from shallow to deep, as shown in Fig.~\ref{fig:DNN_sx_l20_l64}b. The PINN-reconstructed scattered wavefield corresponding to this network is shown in Fig.~\ref{fig:sigsbee_du_ml}d, and we see that it resembles the true scattered wavefield. Fig.~\ref{fig:sigsbee_du_dif}d shows the scattered wavefield difference corresponding to the network with decreasing neurons in each hidden layer, and we observe small differences in the vertical ellipse. In this case, the runtime per epoch using the Adam optimizer is 0.132 s, which is computationally cheaper than using 40 and 60 neurons in each hidden layer of NNs. Though we observe that the reconstructed wavefields in Figs.~\ref{fig:sigsbee_du_ml} are different from the true one in Fig.~\ref{fig:sigsbee_du} in the places far from the source location, these errors will be compensated by the reconstructed wavefields from other sources.

Let us compare the runtimes and scattered wavefield differences for different NN architectures shown in Table 1. We can see that by using the same number of neurons in each hidden layer, the more neurons in the hidden, the smaller error in the scattered wavefield. However, the larger the network, the more runtime is required. Additionally, an unusually large number of neurons in the hidden layers will cause overfitting. Compared to the uniform NNs, a hierarchically decreasing NN is efficient and accurate.

\begin{figure}
\begin{center}
\includegraphics[width=0.5\textwidth]{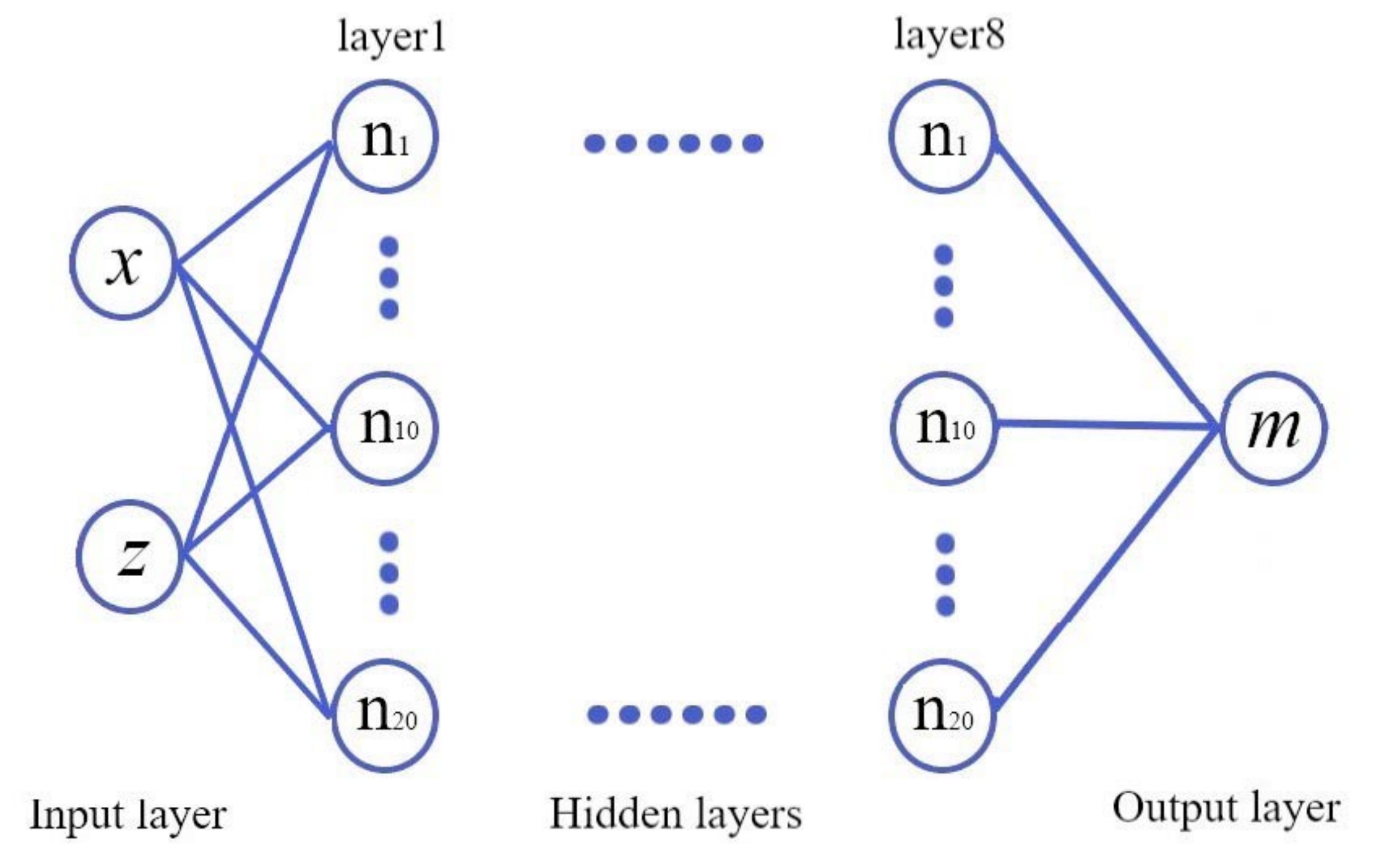} 
\caption{The NN architecture used to predict the velocity solution.}
\label{fig:DNN_v_pinn_n20}
\end{center}   
\end{figure}

\begin{figure}
\begin{center}
\includegraphics[width=1.0\textwidth]{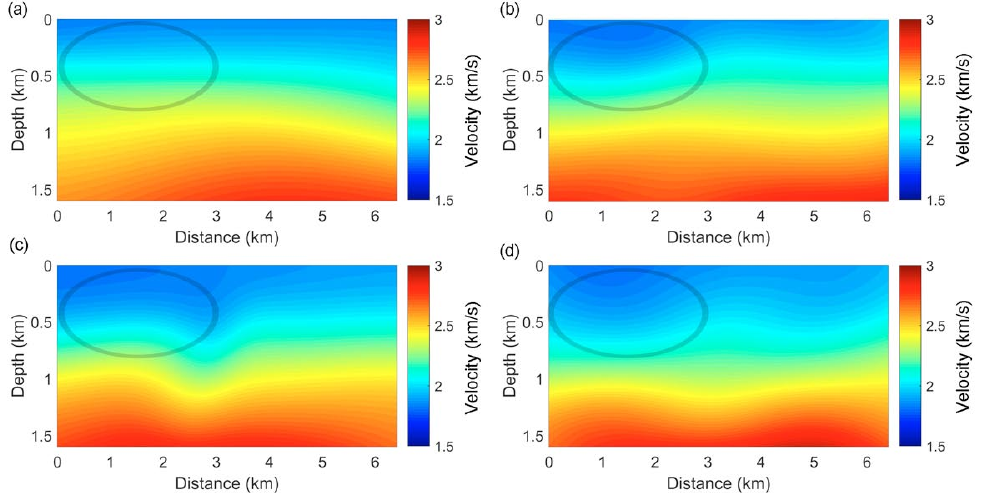} 
\caption{The PINN-inverted velocity models corresponding to the PINN-predicted wavefield in (a) Fig.~\ref{fig:sigsbee_du_ml}a, (b) Fig.~\ref{fig:sigsbee_du_ml}b, (c) Fig.~\ref{fig:sigsbee_du_ml}c, and (d) Fig.~\ref{fig:sigsbee_du_ml}d. The horizontal ellipse regions show low-velocity zone in the shallow part of PINN-based WRI inverted velocity models.}
\label{fig:sigsbee_inv_v}
\end{center}   
\end{figure}

After reconstructing the scattered wavefields using PINN, we build another PINN to invert for the velocity. In this network, we use a 5-layer fully connected NN to predict the velocity, and each hidden layer has 20 neurons, as shown in Fig.~\ref{fig:DNN_v_pinn_n20}. The MSE loss for this NN is given by Eq. 8, and it uses the reconstructed scattered wavefields from all the sources. We use the same activation function and optimizers for this network as the wavefield reconstruction one. As this NN for predicting the velocity is rather small and only first-order spatial derivatives are involved, the computational cost for training this network is quite low. Only 200 epochs of the Adam optimizer is needed to complete the training. We input the regular grid points into the trained network to obtain the predicted velocity model. We show the PINN-predicted velocity models using the PINN-reconstructed scattered wavefields from different NN architectures in Figs.~\ref{fig:sigsbee_inv_v}a-\ref{fig:sigsbee_inv_v}d, respectively. Using the PINN-reconstructed scattered wavefields from the network with 20 neurons in each hidden layer, the inverted velocity fails to capture the low-velocity zone in the horizontal ellipse, as shown in Fig.~\ref{fig:sigsbee_inv_v}a. On the other hand, the inverted velocity models using scattered wavefields reconstructed from larger uniform NN are able to recover the low-velocity zone, as shown in Figs.~\ref{fig:sigsbee_inv_v}b and \ref{fig:sigsbee_inv_v}c, respectively. Fig.~\ref{fig:sigsbee_inv_v}d shows the inverted velocity using scattered wavefields reconstructed from the hierarchically decreasing NN. We show three velocity profiles at locations 0.8 km, 2.4 km, and 4.0 km, respectively, for a detailed comparison. We can see that the inverted velocity using the PINN-reconstructed scattered wavefields from the 20-neuron uniform network does not recover any feature of the true velocity. While the inverted velocity models corresponding to the scattered wavefields from the 20-, 40-neurons uniform and hierarchically decreasing NNs are able to capture the main feature of the true velocity to varying degrees. As we showed above, the hierarchically decreasing NN does the best job in reconstructing the scattered wavefields, as a result, the inverted velocity is closer to the true velocity. We show the $l_{2}$ norms of velocity difference between the true and PINN-based WRI inverted velocity models in Table 1. The $l_{2}$ norm velocity differences corresponding to different NN architectures are consistent with the $l_{2}$ norm scattered wavefield differences, as the reconstructed scattered wavefields control the velocity inversion accuracy.

\begin{figure}
\begin{center}
\includegraphics[width=1.0\textwidth]{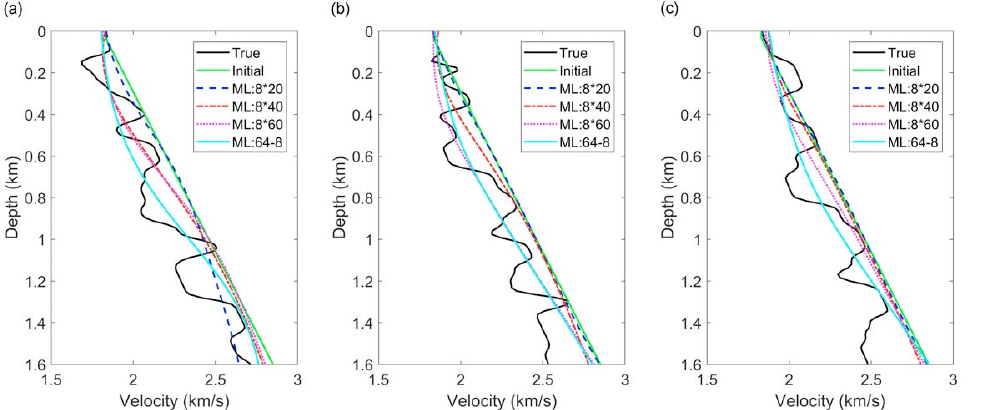} 
\caption{The velocity profiles at locations (a) 0.8 km, (b) 2.4 km, and (c) 4.0 km in Fig.~\ref{fig:sigsbee_inv_v} (black solid curve: true velocity; green solid curve: initial velocity; blue dashed curve: inverted velocity in Fig.~\ref{fig:sigsbee_inv_v}a; red dashed curve: inverted velocity in Fig.~\ref{fig:sigsbee_inv_v}b; purple dotted curve: inverted velocity in Fig.~\ref{fig:sigsbee_inv_v}c; cyan solid curve: inverted velocity in Fig.~\ref{fig:sigsbee_inv_v}d).}
\label{fig:sigsbee_v_profile}
\end{center}   
\end{figure}

\begin{table}[!htbp]
\centering
\caption{Comparisons between different NN architectures}\label{tab:aStrangeTable}%添加标题 设置标签
\begin{tabular}{cccc}
\toprule
NN architecture & Runtime/epoch (s) & $\delta u$ difference &$v$ difference\\
\midrule
20 $\times$ 8 & 0.113& 23.53&  28.53\\
40 $\times$ 8 & 0.151& 20.57& 24.94\\
60 $\times$ 8 & 0.213& 17.61& 20.96\\
64$-$8        & 0.132& 16.94& 18.36\\
\bottomrule
\end{tabular}
%\caption{这是一张三线表}\label{tab:aStrangeTable}  
\end{table}

\begin{figure}
\begin{center}
\includegraphics[width=0.5\textwidth]{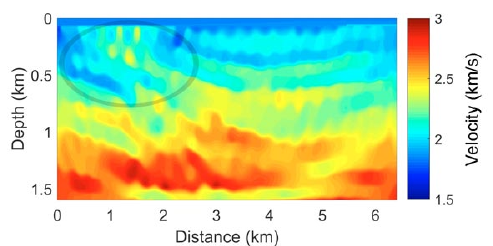} 
\caption{The FWI inverted velocity model using the initial velocity in Fig.~\ref{fig:sigsbee_v_v0}b. The horizontal ellipse region shows the failure of low-velocity zone recovery in the shallow part.}
\label{fig:sigsbee_inv_v0_fwi}
\end{center}   
\end{figure}

\begin{figure}
\begin{center}
\includegraphics[width=1.0\textwidth]{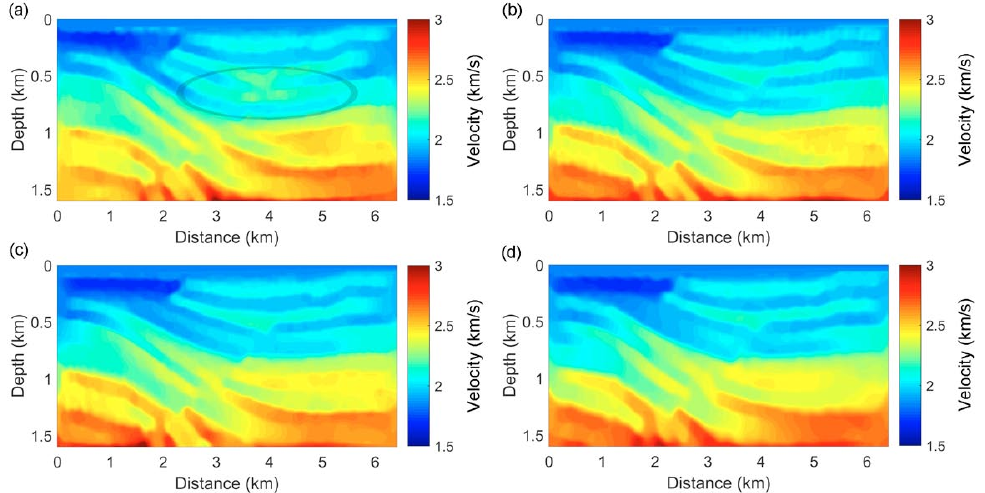} 
\caption{The sequential FWI inverted velocity models using the PINN-predicted velocity models in (a) Fig.~\ref{fig:sigsbee_inv_v}a, (b) Fig.~\ref{fig:sigsbee_inv_v}b, (c) Fig.~\ref{fig:sigsbee_inv_v}c, and (d) Fig.~\ref{fig:sigsbee_inv_v}d. The horizontal ellipse region in Fig.~\ref{fig:sigsbee_inv_v_fwi}a shows the poor recovery in the inverted velocity.}
\label{fig:sigsbee_inv_v_fwi}
\end{center}   
\end{figure}

To evaluate the accuracy of the PINN-based WRI inverted velocity models, we apply the conventional FWI to them. First, we use the initial velocity in Fig.~\ref{fig:sigsbee_v_v0}b to perform FWI. The frequency band we use here is from 4 Hz to 8 Hz with a sampling interval of 0.5 Hz. We perform the frequency-domain FWI from low to high frequency and repeat the full frequency band sweep using 10 outer iterations. A total-variation (TV) regularization is used to smooth the velocity update and preserve the sharp edges \cite{song2019tv}. The inverted velocity using the initial velocity is shown in Fig.~\ref{fig:sigsbee_inv_v0_fwi}, and it fails to capture the low-velocity zone in the horizontal ellipse. Then, we use the PINN-based WRI inverted velocity models as the initial models to perform FWI using the same inversion setup. The sequential FWI inverted velocity models are shown in Figs.~\ref{fig:sigsbee_inv_v_fwi}a-\ref{fig:sigsbee_inv_v_fwi}d, respectively. We observe that the FWI inverted velocity using the initial velocity in Fig.~\ref{fig:sigsbee_inv_v}a is not well recovered in the area circled by a horizontal ellipse. By comparison, PINN-predicted velocity models in Figs.~\ref{fig:sigsbee_inv_v}b-\ref{fig:sigsbee_inv_v}d can be used as good starting models for FWI to get the final inversion results. The results show that PINN-based WRI is able to provide a good initial velocity for FWI with only one iteration and one single frequency for this example.

\begin{figure}
\begin{center}
\includegraphics[width=0.5\textwidth]{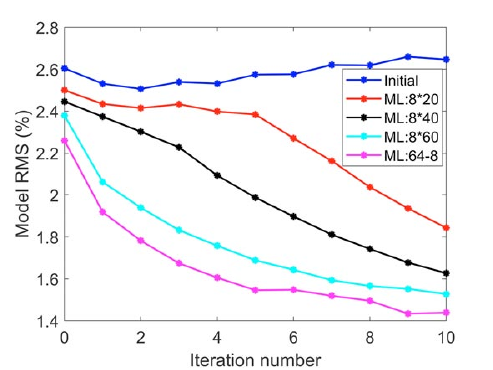} 
\caption{The percent RMS error of the inverted velocity models per iteration for the Sigsbee model. Iteration 0 indicates the RMS errors of starting models for different methods (blue curve: FWI starting from Fig.~\ref{fig:sigsbee_v_v0}b; red curve: FWI starting from Fig.~\ref{fig:sigsbee_inv_v}a; black curve: FWI starting from Fig.~\ref{fig:sigsbee_inv_v}b; cyan curve: FWI starting from Fig.~\ref{fig:sigsbee_inv_v}c; purple curve: FWI starting from Fig.~\ref{fig:sigsbee_inv_v}d).}
\label{fig:sigsbee_v_fwi_rms}
\end{center}   
\end{figure}

To quantitatively demonstrate the quality of starting models for FWI, we show in Fig.~\ref{fig:sigsbee_v_fwi_rms} the percent root-mean-square (RMS) errors in the inverted velocity models per iteration measured with respect to the true model. In Fig.~\ref{fig:sigsbee_v_fwi_rms}, iteration 0 indicates RMS errors of FWI starting models obtained from different NN architectures. It is obvious that FWI does not converge to the true model due to cycle skipping using the raw initial velocity model in Fig.~\ref{fig:sigsbee_v_v0}b. The RMS model errors for PINN-based WRI inverted models are lower than that from the raw initial velocity model in Fig.~\ref{fig:sigsbee_v_v0}b, which demonstrates the velocity improvements by PINN-based WRI. The increase of the neuron number in each hidden layer for uniform NNs will improve the velocity inversion convergence and decrease the RMS model errors. On the other hand, the hierarchically decreasing NN can further improve the inversion result and reach the lowest RMS model error.

\subsection{2D Marmousi model}

\begin{figure}
\begin{center}
\includegraphics[width=1.0\textwidth]{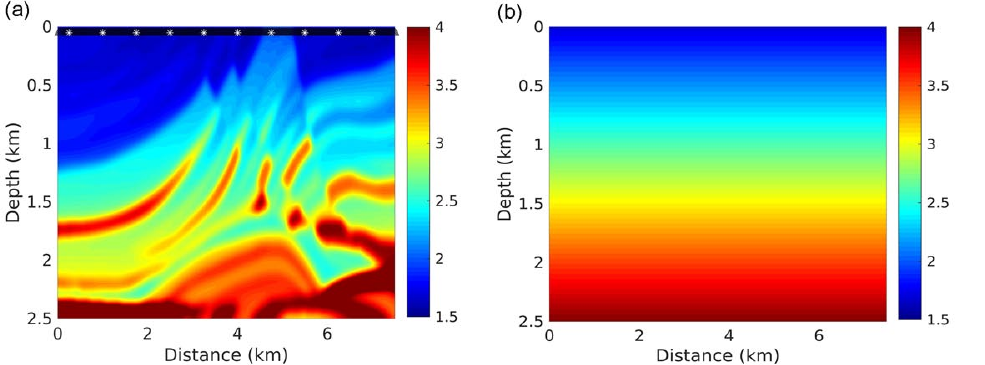} 
\caption{(a) The true modified Marmousi model, and (b) the initial velocity model for inversion implementation.}
\label{fig:mar_true_ini}
\end{center}   
\end{figure}

\begin{figure}
\begin{center}
\includegraphics[width=1.0\textwidth]{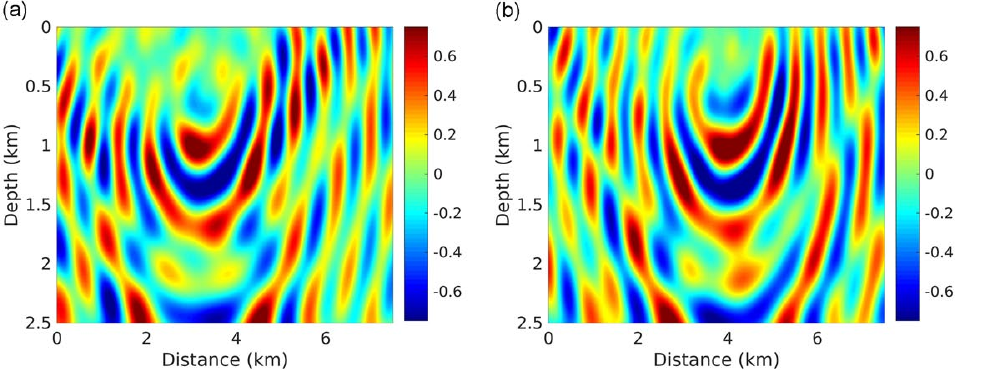} 
\caption{The real part of the scattered wavefields from the numerical method for the (a) third source and (b) sixth source.}
\label{fig:mar_du_true}
\end{center}   
\end{figure}

Next, we apply the proposed PINN-based WRI to a modified Marmousi model and the model is slightly smoothed, as shown in Fig.~\ref{fig:mar_true_ini}a. The initial velocity is linearly increasing with depth, which is shown in Fig.~\ref{fig:mar_true_ini}b. The size of the model is $301 \times 101$ with a grid interval of 25 m in both directions. We place ten sources evenly distributed on the surface, as the $*$ signs in Fig.~\ref{fig:mar_true_ini}a indicate. The data are recorded at a depth of 25 m using 301 receivers. The homogeneous velocity model used to get the background wavefield is 1.7 km/s. We first show the real part of the 3 Hz scattered wavefields using the numerical method corresponding to the third and sixth sources in Figs.~\ref{fig:mar_du_true}a and~\ref{fig:mar_du_true}b, respectively. 

\begin{figure}
\begin{center}
\includegraphics[width=0.5\textwidth]{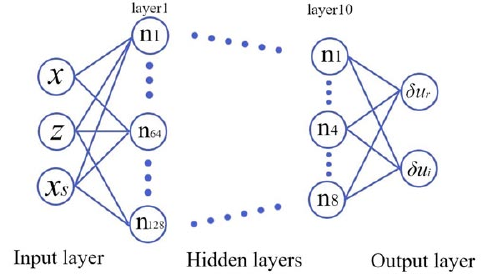} 
\caption{The NN architecture used to reconstruct the scattered wavefield solution.}
\label{fig:DNN_aug_l8n20}
\end{center}   
\end{figure} 

\begin{figure}
\begin{center}
\includegraphics[width=0.5\textwidth]{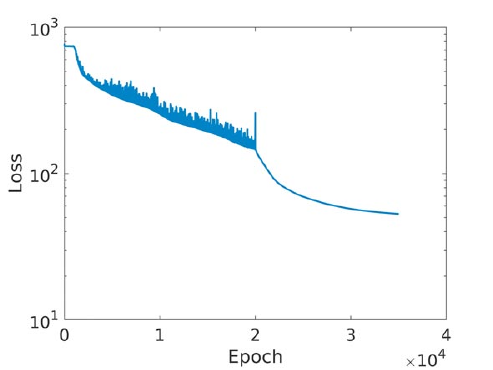} 
\caption{The training loss history for reconstructing the scattered wavefields.}
\label{fig:misfit_du}
\end{center}   
\end{figure}

As the Marmousi model in this example is more complex than the Sigsbee model, we increase the number of hidden layers to 10. We use a hierarchically decreasing NN, which proved to be efficient and accurate in the previous example. The neuron number in each hidden layer is $\left \{128, 128, 64, 64, 32, 32, 16, 16, 8, 8 \right \}$ from shallow to deep, as shown in Fig.~\ref{fig:DNN_aug_l8n20}. We generate 80000 random sets of $\left \{x_{i},z_{i}, x_{si}\right \}$ and feed them into the network. Their corresponding analytically calculated background wavefields and linear interpolated initial velocity (interpolated the random training sample locations) are used in the loss function to train the network. We train the network for 20000 epochs of the Adam optimizer and 15000 iterations of LBFGS, and the loss history curve is shown in Fig.~\ref{fig:misfit_du}. It is obvious that we achieve convergence in the training. We observe that LBFGS have a quicker training loss decrease than the Adam optimizer. However, LBFGS needs to be used after a number of the Adam optimizer training.

\begin{figure}
\begin{center}
\includegraphics[width=1.0\textwidth]{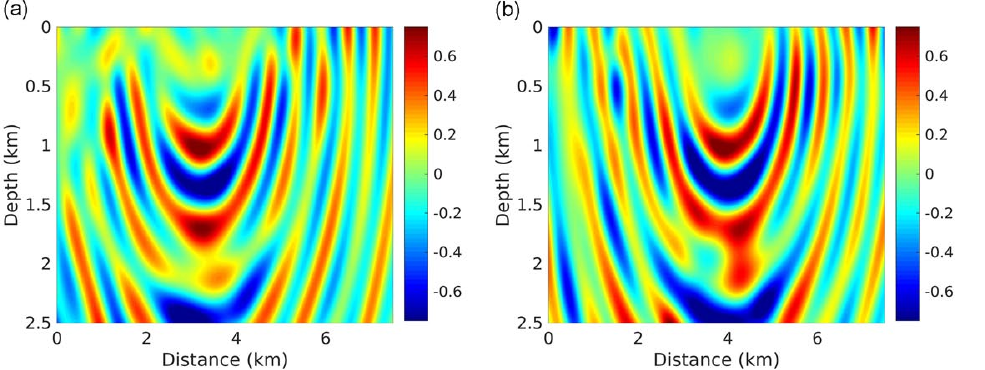} 
\caption{The real part of the scattered wavefields from the PINN method for the (a) third source and (b) sixth source.}
\label{fig:mar_du_ml}
\end{center}   
\end{figure}

\begin{figure}
\begin{center}
\includegraphics[width=1.0\textwidth]{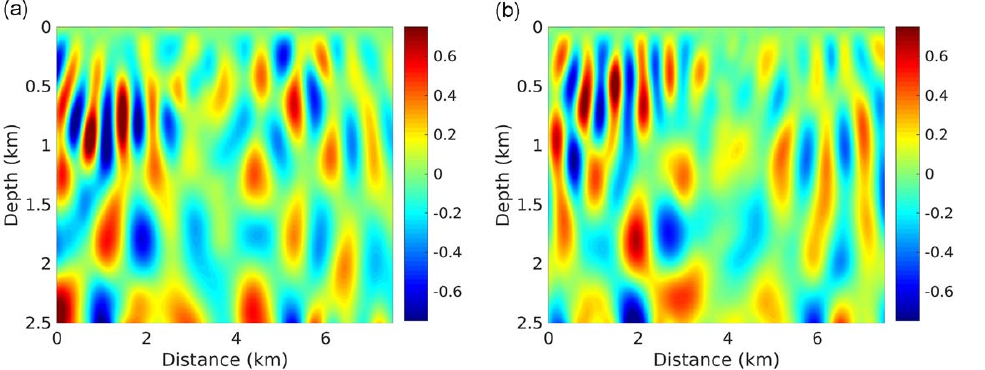} 
\caption{The real part of the scattered wavefield difference between the numerical and PINN methods for the (a) third source and (b) sixth source.}
\label{fig:mar_du_dif}
\end{center}   
\end{figure}

Using the regular grid points as input to the NN, the real parts of the resulting PINN-reconstructed scattered wavefields for the initial velocity (Fig.~\ref{fig:mar_true_ini}b) corresponding to the third and sixth sources are shown in Figs.~\ref{fig:mar_du_ml}a and~\ref{fig:mar_du_ml}b, respectively. It is obvious that the predicted scattered wavefields are generally smooth. We show the real part of the scattered wavefield difference between the numerical and PINN solutions for the same sources in Fig.~\ref{fig:mar_du_dif}. We can observe differences in most areas of the model, and these differences are smaller up shallow, where the receivers reside. The PINN-reconstructed wavefields are generally smooth and lack the detailed information corresponding to the small scatterers. These smooth reconstructed wavefields are helpful in building smooth initial velocity models.

\begin{figure}
\begin{center}
\includegraphics[width=0.5\textwidth]{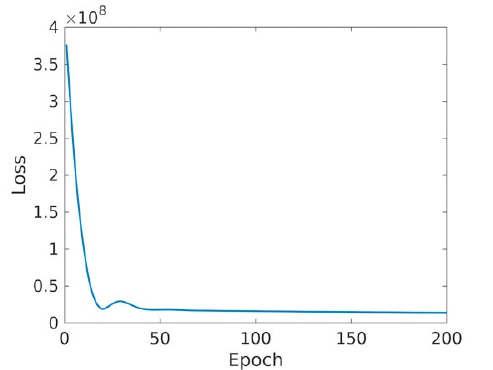} 
\caption{The training loss history for predicting the velocity.}
\label{fig:misfit_v}
\end{center}   
\end{figure}

\begin{figure}
\begin{center}
\includegraphics[width=1.0\textwidth]{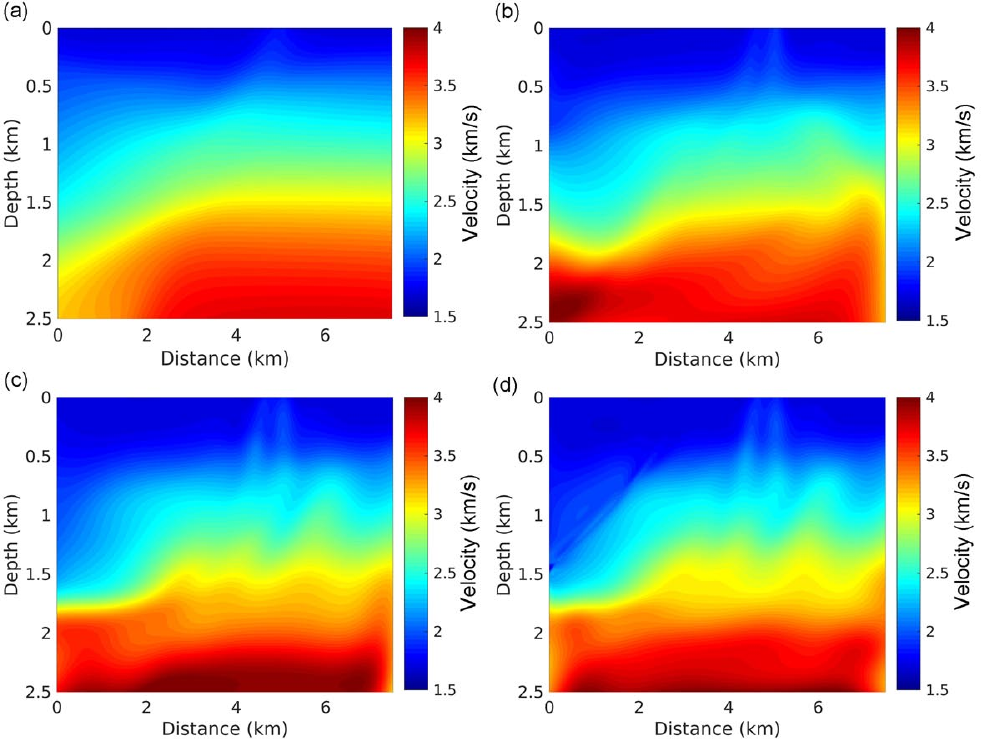} 
\caption{The PINN-based WRI inverted velocity models (a) after one iteration using 3 Hz, (b) after four iterations using 3 Hz, (c) after four iterations using 4 Hz, and (d) after three iterations using 5 Hz.}
\label{fig:mar_inv}
\end{center}   
\end{figure}

After obtaining the PINN-reconstructed scattered wavefields, we use the network in Fig.~\ref{fig:DNN_v_pinn_n20} to invert for the velocity. We train this network using the same velocity prediction training setup as the previous example. The training loss history for predicting the velocity is shown in Fig.~\ref{fig:misfit_v}. We can see that the training converges very fast because the velocity model is a lot smoother than the wavefield. For this model, we use three frequencies which are 3, 4, 5 Hz, to ultimately invert for the velocity. Following the classic waveform inversion implementation, we start the inversion process from the low frequency. After the first iteration, the inverted velocity corresponding to the PINN-reconstructed scattered wavefields is shown in Fig.~\ref{fig:mar_inv}a. It is obvious that the inverted velocity is very smooth and reflects the general structure of the true model. After another three iterations (four iterations in total) using 3 Hz, the inverted velocity is shown in Fig.~\ref{fig:mar_inv}b. We see more structural information of the true velocity is recovered. Then, we move to the next frequency, which is 4 Hz. After 4 iterations using 4 Hz, we obtain the inverted velocity shown in Fig.~\ref{fig:mar_inv}c. From the higher frequency, we obtain additional, higher resolution, information, but the model is still generally smooth. Finally, we perform three iterations using 5 Hz, and obtain the velocity shown in Fig.~\ref{fig:mar_inv}d. As expected, we get additional, yet mild, details in the velocity. This is because PINN provided us with smooth scattered wavefields, which result in a smooth inverted velocity. However, this PINN-based WRI inverted velocity is a good initial model for a further FWI application.

\begin{figure}
\begin{center}
\includegraphics[width=1.0\textwidth]{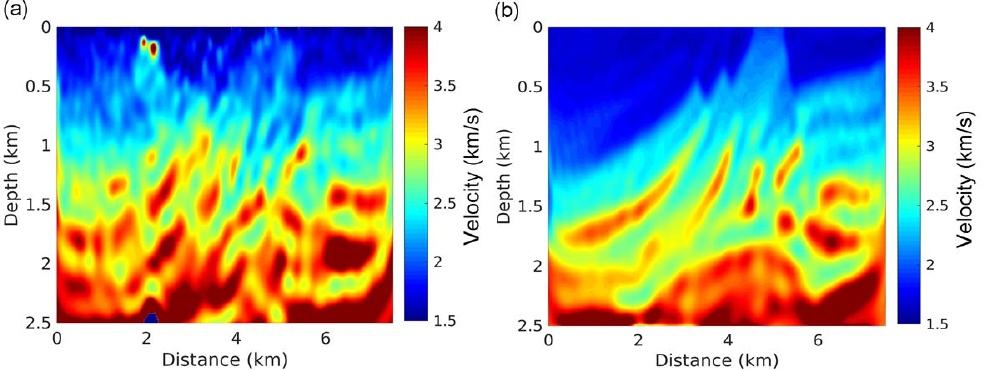} 
\caption{(a) The FWI inverted velocity using the initial velocity in Fig.~\ref{fig:mar_true_ini}b, and (b) the FWI inverted velocity using the PINN-based WRI inverted velocity Fig.~\ref{fig:mar_inv}d.}
\label{fig:mar_fwi}
\end{center}   
\end{figure}

\begin{figure}
\begin{center}
\includegraphics[width=1.0\textwidth]{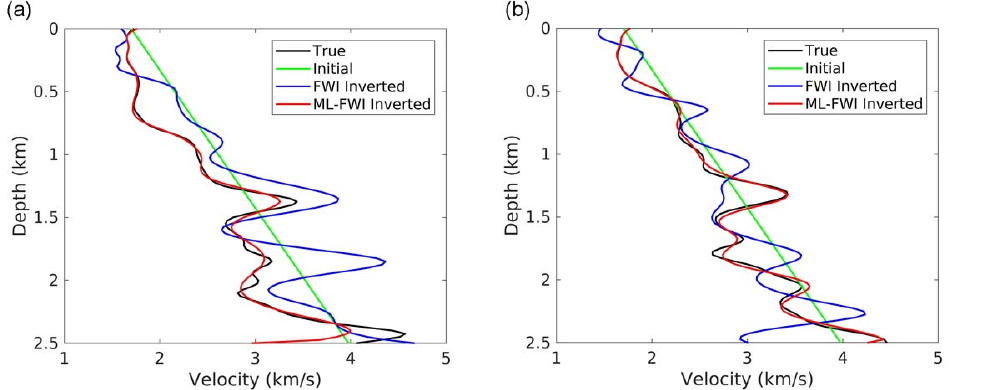} 
\caption{The vertical velocity profiles at locations (a) 2.5 km and (b) 3.75 km in Fig.~\ref{fig:mar_true_ini} (black curve: true velocity; green curve: initial velocity; blue curve: FWI inverted velocity from Fig.~\ref{fig:mar_true_ini}b; red curve: FWI inverted velocity from Fig.~\ref{fig:mar_fwi}d).}
\label{fig:mar_profiles}
\end{center}   
\end{figure}

Again, we apply the conventional FWI to the inverted velocity in Fig.~\ref{fig:mar_inv}d. First, we use the initial velocity (Fig.~\ref{fig:mar_true_ini}b) to perform the FWI. The frequency band we use here is from 3 Hz to 7 Hz with a sampling interval of 1 Hz. We perform 50 iterations to update the velocity, and the inverted velocity model is shown in Fig.~\ref{fig:mar_fwi}a. It is obvious that FWI suffers from cycle skipping, as the initial velocity is crude. Then, we repeat the FWI implementation with the same inversion setup using the PINN-based WRI inverted velocity in Fig.~\ref{fig:mar_inv}d as a starting model, and the resulting FWI-inverted velocity is shown in Fig.~\ref{fig:mar_fwi}b. Clearly, the sequential FWI inverted velocity after PINN-based WRI is able to recover the details in the true velocity. We show two vertical velocity profiles at locations 2.5 km and 3.75 km in Figs.~\ref{fig:mar_profiles}a and~\ref{fig:mar_profiles}b, respectively. In both vertical velocity profile comparisons, we can see that the sequential FWI inverted velocity after the PINN-based WRI has a much better agreement with the true velocity than that from the direct FWI implementation.

\section{Discussions}

In PINN-based WRI implementation, we tend to reconstruct the scattered wavefield instead of the whole wavefield to mitigate the point source singularity, which will hamper the training convergence. By using the scattered wave equation, the source term is related to the velocity perturbation and background wavefield. Thus, it is extended from a point to the whole domain. We use an infinite isotropic homogeneous velocity to analytically obtain the background wavefield. We recommend using a velocity given by the source location velocity so the background wavefield would absorb the majority of the source singularity effect. If the source is located in the water layer, like in a marine survey, we can just use the water velocity to generate the background wavefields. In this case, the target scattered wavefields do not have especially large values near the source location. For each source, a number of random points, which is far fewer than the regular grid points, are selected within the domain of interest to train the network used to reconstruct the scattered wavefields.

We use two independent NNs to reconstruct the scattered wavefields and predict the velocity. In the network to reconstruct the scattered wavefields, with the same number of hidden layers, an increase of the number of neurons will improve the scattered wavefield reconstruction accuracy. However, an unpractical large number of neurons in the network will cause overfitting and increase the computational cost \cite{hawkins2004problem}. Based on the experiments we performed, we found that an NN with a decreasing number of neurons from shallow to deep layers shows better results than using the same number of neurons in all the hidden layers \cite{mhaskar2016learning,giannakis2019machine,gao2019optimized}. By doing so, the first few large layers can learn many low-level features, and these lower-level features will be fed into the subsequent smaller layers. As a result, higher-order features can be extracted efficiently and sufficiently by deep thinner layers. For the first simple example, we use an 8-layer fully connected NN to reconstruct the wavefields, and this PINN architecture is borrowed from previous work \cite{helmholtz,alkhalifah2020machine}, which are proven to be effective. For the second Marmousi model, as the complexity of the velocity increases, a deeper network and more neurons in the hidden layers is required to train the network. The NN used to predict the velocity model is much smaller than the one used to reconstruct the wavefield. This is because wavefield-reconstruction NN requires three inputs, and it needs to evaluate second-order spatial derivatives. While the velocity prediction NN just requires two inputs, and only first-order spatial derivatives are needed for the TV regularization.

We used one single NN to reconstruct the scattered wavefields for all the sources set in the seismic survey by introducing the source horizontal location as input. This network will save a tremendous computational cost compared to reconstructing scattered wavefields using an individual network for each source. However, the computational cost in the PINN training is still overall larger than calculating the matrix inverse using the FD method for small 2D models of isotropic acoustic media. This is the main limitation of the proposed method. Conservative PINN (cPINN) partitions the computational domain into several subdomains to enable easy parallelization and fast convergence \cite{jagtap2020conservative}. We will try to incorporate the idea of cPINN in future work.

The proposed PINN-based WRI may provide a feasible solution to 3D large models in complex media thanks to the development in the GPU hardware and parallel training implementation. On the other hand, the computational cost for calculating the inverse of the impedance matrix increases exponentially for the FD method when the model size increases. In addition, the impedance matrix for acoustic VTI media and isotropic elastic media will be twice as large as the case for isotropic media in both vertical and horizontal directions. As a result, the computational cost for solving the acoustic VTI and isotropic elastic wave equations will be 8 times that of the acoustic wave equation \cite{wu2018efficient}. Additionally, the size of the impedance matrix will be even larger for more complex media, like viscoelastic anisotropic \cite{yang2020multiparameter} and fluid-saturated porous media \cite{yang2020new}. PINN has the potential in improving the computational efficiency of waveform inversion in media corresponding to complex physics. Another limitation in this paper is that we currently focus on 2D models. For 3D models, the input data to the network should be the 3D spatial coordinate values $\left \{x,y,z \right \}$ and source locations $\left \{x_{s},y_{s}\right \}$ on the surface. Thus, a new NN architecture needs to be studied. PINN-based WRI has excellent flexibility and versatility for different media and model dimensions as we only need to modify the loss functions corresponding to the different wave equations. By comparison, we need tremendous modifications to convert the conventional WRI from isotropic media to anisotropic media, and also from 2D models to 3D models. The applications of the proposed method on 3D models and complex media remain to be investigated.

Both WRI and EWI are able to provide high-resolution velocity updates, as they utilize the reflection wave information. The key step of these two methods is to reconstruct accurate wavefields. However, the proposed PINN-based method can only reconstruct generally smooth scattered wavefields and result in a smooth inverted velocity model. Even though we increase the frequency used in the PINN-based WRI, it still fails to provide high-resolution components to the inverted velocity due to the smoothness of the reconstructed scattered wavefields. This apparent limitation is the focus of our current research efforts. We tend to resolve this issue using a periodic activation function in a paper we have under review \cite{sitzmann2020implicit}. On the other hand, the smooth inverted velocity from PINN-based WRI can be used as a good initial velocity for further FWI application, and we obtain the inversion result using very limited iterations and frequencies. The weighting factor $\alpha$ in Eq. 8 plays an important role in reconstructing the scattered wavefield. It is used to balance the data constraint and the physics constraint, so the value of $\alpha$ should equalize the magnitudes of these two terms. Trial and error tests are required to find an optimal weighting factor.

Unlike the resulting vorticity from the Navier–Stokes equation, which is quite smooth \cite{raissi2019physics,li2020fourier}, the high-frequency wavefield has obvious sinusoidal features. Thus, it is a larger challenge for PINN to express the complexity of a high-frequency wavefield. The proposed PINN-based WRI has a limitation for high-frequency wavefields reconstruction. In the Marmousi example, we show that multiple high-frequency components will not refine too many details of the inverted velocity model. Thus, we recommend using the lowest frequency component available in the data to perform the proposed method and retrieve the background velocity model. Finally, a sequential FWI implementation provides us an accurate high-resolution inverted velocity model.

\section{Conclusion}

We proposed a physics-informed neural network (PINN) based wavefield reconstruction inversion (WRI) method. By using two independent PINNs, we can solve the augmented wave equation and estimate the velocity, sequentially. We combined the scattered wave equation and data-fitting terms as the loss function to reconstruct the scattered wavefield in the first network. Then, we used the reconstructed scattered wavefield to predict the velocity using the scattered wave equation as the loss function. We analyzed how NN architectures affect the efficiency and accuracy of the proposed method and showed the superiority of a hierarchically decreasing NN. Applications on a part of the Sigsbee2A model and a modified Marmousi model show that the PINN-based WRI can invert for a reasonably smooth velocity with very limited iterations and frequencies, which can be a good initial velocity for FWI.

\section*{Acknowledgment}

We thank KAUST for its support and the SWAG group for the collaborative environment. This work utilized the resources of the Supercomputing Laboratory at King Abdullah University of Science and Technology (KAUST) in Thuwal, Saudi Arabia, and we are grateful for that.

\bibliographystyle{unsrt}  
\bibliography{refs/pinn_helm_vti}

\begin{thebibliography}{10}

\bibitem{tarantola}
Albert Tarantola.
\newblock Inversion of seismic reflection data in the acoustic approximation.
\newblock {\em Geophysics}, 49(8):1259--1266, 1984.

\bibitem{vir}
Jean Virieux and St{\'e}phane Operto.
\newblock An overview of full-waveform inversion in exploration geophysics.
\newblock {\em Geophysics}, 74(6):WCC1--WCC26, 2009.

\bibitem{clement}
F.~Clement, G.~Chavent, and Susana G.
\newblock Migration-based traveltime waveform inversion of 2-d simple
  structures: A synthetic example.
\newblock {\em Geophysics}, 66:845--860, 2001.

\bibitem{biondi}
Biondo Biondi and W.~Symes.
\newblock Angle-domain common-image gathers for migration velocity analysis by
  wavefield-continuation imaging.
\newblock {\em Geophysics}, 69:1283--1298, 2004.

\bibitem{symes}
William~W Symes.
\newblock Migration velocity analysis and waveform inversion.
\newblock {\em Geophysical prospecting}, 56:765--790, 2008.

\bibitem{chi2014full}
Benxin Chi, Liangguo Dong, and Yuzhu Liu.
\newblock Full waveform inversion method using envelope objective function
  without low frequency data.
\newblock {\em Journal of Applied Geophysics}, 109:36--46, 2014.

\bibitem{wu2014seismic}
Ru-Shan Wu, Jingrui Luo, and Bangyu Wu.
\newblock Seismic envelope inversion and modulation signal model.
\newblock {\em Geophysics}, 79(3):WA13--WA24, 2014.

\bibitem{chen2020application}
Guoxin Chen, Wencai Yang, Shengchang Chen, Yanan Liu, and Zhiwei Gu.
\newblock Application of envelope in salt structure velocity building: from
  objective function construction to the full-band seismic data reconstruction.
\newblock {\em IEEE Transactions on Geoscience and Remote Sensing},
  58(9):6594--6608, 2020.

\bibitem{bozdaug2011misfit}
Ebru Bozda{\u{g}}, Jeannot Trampert, and Jeroen Tromp.
\newblock Misfit functions for full waveform inversion based on instantaneous
  phase and envelope measurements.
\newblock {\em Geophysical Journal International}, 185(2):845--870, 2011.

\bibitem{choi2013frequency}
Yunseok Choi and Tariq Alkhalifah.
\newblock Frequency-domain waveform inversion using the phase derivative.
\newblock {\em Geophysical Journal International}, 195(3):1904--1916, 2013.

\bibitem{xu}
Su~Xu, D.~Wang, F.~Chen, and Y.~Zhang.
\newblock Full waveform inversion for reflected seismic data.
\newblock {\em 74th Annual International Conference and Exhibition, EAGE,
  Extend Abstracts}, page W024, 2012.

\bibitem{zhou}
Hongbo Zhou, L.~Amundsen, and G.~Zhang.
\newblock Fundamental issues in full waveform inversion.
\newblock {\em 82nd Annual International Meeting, SEG, Expanded Abstracts},
  pages 1--5, 2012.

\bibitem{wang2020enhancing}
Guanchao Wang, Tariq Alkhalifah, and Shangxu Wang.
\newblock Enhancing low-wavenumber information in reflection waveform inversion
  by the energy norm born scattering.
\newblock {\em IEEE Geoscience and Remote Sensing Letters}, 2020.

\bibitem{song2020reflection}
Chao Song and Tariq Alkhalifah.
\newblock A reflection-based efficient wavefield inversion.
\newblock {\em Geophysics}, 2021.

\bibitem{warner2016adaptive}
Michael Warner and Llu{\'\i}s Guasch.
\newblock Adaptive waveform inversion: Theory.
\newblock {\em Geophysics}, 81(6):R429--R445, 2016.

\bibitem{sun2020joint}
Bingbing Sun and Tariq~A Alkhalifah.
\newblock Joint minimization of the mean and information entropy of the
  matching filter distribution for a robust misfit function in full-waveform
  inversion.
\newblock {\em IEEE Transactions on Geoscience and Remote Sensing},
  58(7):4704--4720, 2020.

\bibitem{van2013}
Tristan van Leeuwen and Felix~J Herrmann.
\newblock Mitigating local minima in full-waveform inversion by expanding the
  search space.
\newblock {\em Geophysical Journal International}, 195(1):661--667, 2013.

\bibitem{van2015}
Tristan van Leeuwen and Felix~J Herrmann.
\newblock A penalty method for pde-constrained optimization in inverse
  problems.
\newblock {\em Inverse Problems}, 32(1):015007, 2015.

\bibitem{alkhalifah2019efficient}
Tariq Alkhalifah and Chao Song.
\newblock An efficient wavefield inversion: Using a modified source function in
  the wave equation.
\newblock {\em Geophysics}, 84(6):R909--R922, 2019.

\bibitem{song2019tv}
Chao Song and Tariq Alkhalifah.
\newblock Efficient wavefield inversion with outer iterations and total
  variation constraint.
\newblock {\em IEEE Transactions on Geoscience and Remote Sensing},
  58(8):5836--5846, 2020.

\bibitem{vapnik2013nature}
Vladimir Vapnik.
\newblock {\em The nature of statistical learning theory}.
\newblock Springer science \& business media, 2013.

\bibitem{zhao2015comparison}
Tao Zhao, Vikram Jayaram, Atish Roy, and Kurt~J Marfurt.
\newblock A comparison of classification techniques for seismic facies
  recognition.
\newblock {\em Interpretation}, 3(4):SAE29--SAE58, 2015.

\bibitem{song2018source}
Chao Song, T~Alkhalifah, and Z~Wu.
\newblock Source type classification based on the support vector machine
  method.
\newblock In {\em 80th EAGE Conference and Exhibition 2018}, volume 2018, pages
  1--5. European Association of Geoscientists \& Engineers, 2018.

\bibitem{li2020separation}
Jing Li, Yuqing Chen, and Gerard~T Schuster.
\newblock Separation of multi-mode surface waves by supervised machine learning
  methods.
\newblock {\em Geophysical Prospecting}, 68(4):1270--1280, 2020.

\bibitem{chen2020suppressing}
Yuqing Chen, Yunsong Huang, and Lianjie Huang.
\newblock Suppressing migration image artifacts using a support vector machine
  method.
\newblock {\em Geophysics}, 85(5):1--55, 2020.

\bibitem{microsvm}
C~Song and T~Alkhalifah.
\newblock Identifying microseismic events in time-reversed source images using
  support vector machine.
\newblock 2020(1):1--5, 2020.

\bibitem{zhu2019phasenet}
Weiqiang Zhu and Gregory~C Beroza.
\newblock Phasenet: a deep-neural-network-based seismic arrival-time picking
  method.
\newblock {\em Geophysical Journal International}, 216(1):261--273, 2019.

\bibitem{kaur2020improving}
Harpreet Kaur, Nam Pham, and Sergey Fomel.
\newblock Improving resolution of migrated images by approximating the inverse
  {H}essian using deep learning.
\newblock {\em Geophysics}, 85(4):1--62, 2020.

\bibitem{zhang2020high}
Zhen-dong Zhang and Tariq Alkhalifah.
\newblock High-resolution reservoir characterization using deep learning-aided
  elastic full-waveform inversion: The north sea field data example.
\newblock {\em Geophysics}, 85(4):WA137--WA146, 2020.

\bibitem{liyy}
Y~Li, T~Alkhalifah, and Z~Zhang.
\newblock High-resolution regularized elastic full waveform inversion assisted
  by deep learning.
\newblock 2020(1):1--5, 2020.

\bibitem{wang2020well}
Yuqing Wang, Qiang Ge, Wenkai Lu, and Xinfei Yan.
\newblock Well-logging constrained seismic inversion based on closed-loop
  convolutional neural network.
\newblock {\em IEEE Transactions on Geoscience and Remote Sensing},
  58(8):5564--5573, 2020.

\bibitem{shi2018automatic}
Yunzhi Shi, Xinming Wu, and Sergey Fomel.
\newblock Automatic salt-body classification using a deep convolutional neural
  network.
\newblock In {\em SEG Technical Program Expanded Abstracts 2018}, pages
  1971--1975. Society of Exploration Geophysicists, 2018.

\bibitem{wu2019faultseg3d}
Xinming Wu, Luming Liang, Yunzhi Shi, and Sergey Fomel.
\newblock Faultseg3d: Using synthetic data sets to train an end-to-end
  convolutional neural network for 3d seismic fault segmentation.
\newblock {\em Geophysics}, 84(3):IM35--IM45, 2019.

\bibitem{ovcharenko2019deep}
Oleg Ovcharenko, Vladimir Kazei, Mahesh Kalita, Daniel Peter, and Tariq
  Alkhalifah.
\newblock Deep learning for low-frequency extrapolation from multioffset
  seismic data.
\newblock {\em Geophysics}, 84(6):R989--R1001, 2019.

\bibitem{moseley2020solving}
Ben Moseley, Tarje Nissen-Meyer, and Andrew Markham.
\newblock Deep learning for fast simulation of seismic waves in complex media.
\newblock {\em Solid Earth}, 11(4):1527--1549, 2020.

\bibitem{yang2019deep}
Fangshu Yang and Jianwei Ma.
\newblock Deep-learning inversion: A next-generation seismic velocity model
  building method.
\newblock {\em Geophysics}, 84(4):R583--R599, 2019.

\bibitem{zhang2020data}
Zhongping Zhang and Youzuo Lin.
\newblock Data-driven seismic waveform inversion: A study on the robustness and
  generalization.
\newblock {\em IEEE Transactions on Geoscience and Remote Sensing},
  58(10):6900--6913, 2020.

\bibitem{kazei2020velocity}
Vladimir Kazei, Oleg Ovcharenko, and Tariq Alkhalifah.
\newblock Velocity model building by deep learning: From general synthetics to
  field data application.
\newblock In {\em SEG Technical Program Expanded Abstracts 2020}, pages
  1561--1565. Society of Exploration Geophysicists, 2020.

\bibitem{ren2020physics}
Yuxiao Ren, Xinji Xu, Senlin Yang, Lichao Nie, and Yangkang Chen.
\newblock A physics-based neural-network way to perform seismic full waveform
  inversion.
\newblock {\em IEEE Access}, 8:112266--112277, 2020.

\bibitem{8931232}
Shucai Li, Bin Liu, Yuxiao Ren, Yangkang Chen, Senlin Yang, Yunhai Wang, and
  Peng Jiang.
\newblock Deep-learning inversion of seismic data.
\newblock {\em IEEE Transactions on Geoscience and Remote Sensing},
  58(3):2135--2149, 2020.

\bibitem{liu2021deep}
Bin Liu, Senlin Yang, Yuxiao Ren, Xinji Xu, Peng Jiang, and Yangkang Chen.
\newblock Deep-learning seismic full-waveform inversion for realistic
  structural models.
\newblock {\em Geophysics}, 86(1):R31--R44, 2021.

\bibitem{giannakis2019machine}
Iraklis Giannakis, Antonios Giannopoulos, and Craig Warren.
\newblock A machine learning-based fast-forward solver for ground penetrating
  radar with application to full-waveform inversion.
\newblock {\em IEEE Transactions on Geoscience and Remote Sensing},
  57(7):4417--4426, 2019.

\bibitem{liu2020deep}
Bin Liu, Qian Guo, Shucai Li, Benchao Liu, Yuxiao Ren, Yonghao Pang, Xu~Guo,
  Lanbo Liu, and Peng Jiang.
\newblock Deep learning inversion of electrical resistivity data.
\newblock {\em IEEE Transactions on Geoscience and Remote Sensing},
  58(8):5715--5728, 2020.

\bibitem{li2018machine}
Zefeng Li, Men-Andrin Meier, Egill Hauksson, Zhongwen Zhan, and Jennifer
  Andrews.
\newblock Machine learning seismic wave discrimination: Application to
  earthquake early warning.
\newblock {\em Geophysical Research Letters}, 45(10):4773--4779, 2018.

\bibitem{jiao2020artificial}
Pengcheng Jiao and Amir~H Alavi.
\newblock Artificial intelligence in seismology: Advent, performance and future
  trends.
\newblock {\em Geoscience Frontiers}, 11(3):739--744, 2020.

\bibitem{rudy2017data}
Samuel~H Rudy, Steven~L Brunton, Joshua~L Proctor, and J~Nathan Kutz.
\newblock Data-driven discovery of partial differential equations.
\newblock {\em Science Advances}, 3(4):e1602614, 2017.

\bibitem{han2018solving}
Jiequn Han, Arnulf Jentzen, and E~Weinan.
\newblock Solving high-dimensional partial differential equations using deep
  learning.
\newblock {\em Proceedings of the National Academy of Sciences},
  115(34):8505--8510, 2018.

\bibitem{sirignano2018dgm}
Justin Sirignano and Konstantinos Spiliopoulos.
\newblock Dgm: A deep learning algorithm for solving partial differential
  equations.
\newblock {\em Journal of computational physics}, 375:1339--1364, 2018.

\bibitem{berg2019data}
Jens Berg and Kaj Nystr{\"o}m.
\newblock Data-driven discovery of pdes in complex datasets.
\newblock {\em Journal of Computational Physics}, 384:239--252, 2019.

\bibitem{tompson2017accelerating}
Jonathan Tompson, Kristofer Schlachter, Pablo Sprechmann, and Ken Perlin.
\newblock Accelerating eulerian fluid simulation with convolutional networks.
\newblock In {\em International Conference on Machine Learning}, pages
  3424--3433. PMLR, 2017.

\bibitem{geneva2020modeling}
Nicholas Geneva and Nicholas Zabaras.
\newblock Modeling the dynamics of pde systems with physics-constrained deep
  auto-regressive networks.
\newblock {\em Journal of Computational Physics}, 403:109056, 2020.

\bibitem{thuerey2020deep}
Nils Thuerey, Konstantin Wei{\ss}enow, Lukas Prantl, and Xiangyu Hu.
\newblock Deep learning methods for reynolds-averaged navier--stokes
  simulations of airfoil flows.
\newblock {\em AIAA Journal}, 58(1):25--36, 2020.

\bibitem{li2020neural}
Zongyi Li, Nikola Kovachki, Kamyar Azizzadenesheli, Burigede Liu, Kaushik
  Bhattacharya, Andrew Stuart, and Anima Anandkumar.
\newblock Neural operator: Graph kernel network for partial differential
  equations.
\newblock {\em arXiv preprint arXiv:2003.03485}, 2020.

\bibitem{li2020fourier}
Zongyi Li, Nikola Kovachki, Kamyar Azizzadenesheli, Burigede Liu, Kaushik
  Bhattacharya, Andrew Stuart, and Anima Anandkumar.
\newblock Fourier neural operator for parametric partial differential
  equations.
\newblock {\em arXiv preprint arXiv:2010.08895}, 2020.

\bibitem{raissi2019physics}
Maziar Raissi, Paris Perdikaris, and George~E Karniadakis.
\newblock Physics-informed neural networks: A deep learning framework for
  solving forward and inverse problems involving nonlinear partial differential
  equations.
\newblock {\em Journal of Computational Physics}, 378:686--707, 2019.

\bibitem{baydin2017automatic}
At{\i}l{\i}m~G{\"u}nes Baydin, Barak~A Pearlmutter, Alexey~Andreyevich Radul,
  and Jeffrey~Mark Siskind.
\newblock Automatic differentiation in machine learning: a survey.
\newblock {\em The Journal of Machine Learning Research}, 18(1):5595--5637,
  2017.

\bibitem{eikonal}
UB~Waheed, Ehsan Haghighat, Tariq Alkhalifah, Chao Song, and Qi~Hao.
\newblock Eikonal solution using physics-informed neural networks.
\newblock In {\em 82nd EAGE Annual Conference \& Exhibition}, volume 2020,
  pages 1--5. European Association of Geoscientists \& Engineers, 2020.

\bibitem{waheed2020anisotropic}
Umair~bin Waheed, Ehsan Haghighat, and Tariq Alkhalifah.
\newblock Anisotropic eikonal solution using physics-informed neural networks.
\newblock In {\em SEG Technical Program Expanded Abstracts 2020}, pages
  1566--1570. Society of Exploration Geophysicists, 2020.

\bibitem{helmholtz}
Tariq Alkhalifah, Chao Song, Q~Hao, et~al.
\newblock Wavefield solutions from machine learned functions that approximately
  satisfy the wave equation.
\newblock In {\em 82nd EAGE Annual Conference \& Exhibition}, volume 2020,
  pages 1--5. European Association of Geoscientists \& Engineers, 2020.

\bibitem{song2020solving}
Chao Song, Tariq Alkhalifah, and Umair~Bin Waheed.
\newblock Solving the frequency-domain acoustic vti wave equation using
  physics-informed neural networks.
\newblock {\em Geophysical Journal International}, 2021.

\bibitem{waheed2021pinntomo}
Umair~bin Waheed, Tariq Alkhalifah, Ehsan Haghighat, Chao Song, and Jean
  Virieux.
\newblock Pinntomo: Seismic tomography using physics-informed neural networks.
\newblock {\em arXiv preprint arXiv:2104.01588}, 2021.

\bibitem{sahli2020physics}
Francisco Sahli~Costabal, Yibo Yang, Paris Perdikaris, Daniel~E Hurtado, and
  Ellen Kuhl.
\newblock Physics-informed neural networks for cardiac activation mapping.
\newblock {\em Frontiers in Physics}, 8:42, 2020.

\bibitem{raissi2020hidden}
Maziar Raissi, Alireza Yazdani, and George~Em Karniadakis.
\newblock Hidden fluid mechanics: Learning velocity and pressure fields from
  flow visualizations.
\newblock {\em Science}, 367(6481):1026--1030, 2020.

\bibitem{alkhalifah2020machine}
Tariq Alkhalifah, Chao Song, and Umair~bin Waheed.
\newblock Machine learned {G}reen's functions that approximately satisfy the
  wave equation.
\newblock In {\em SEG Technical Program Expanded Abstracts 2020}, pages
  2638--2642. Society of Exploration Geophysicists, 2020.

\bibitem{lippmann1950variational}
Bernard~A Lippmann and Julian Schwinger.
\newblock Variational principles for scattering processes. i.
\newblock {\em Physical Review}, 79(3):469, 1950.

\bibitem{engquist2018approximate}
Bj{\"o}rn Engquist and Hongkai Zhao.
\newblock Approximate separability of the green's function of the helmholtz
  equation in the high frequency limit.
\newblock {\em Communications on Pure and Applied Mathematics},
  71(11):2220--2274, 2018.

\bibitem{kingma2014adam}
Diederik~P Kingma and Jimmy Ba.
\newblock Adam: A method for stochastic optimization.
\newblock {\em arXiv preprint arXiv:1412.6980}, 2014.

\bibitem{liu1989limited}
Dong~C Liu and Jorge Nocedal.
\newblock On the limited memory bfgs method for large scale optimization.
\newblock {\em Mathematical programming}, 45(1-3):503--528, 1989.

\bibitem{jo1996optimal}
Churl-Hyun Jo, Changsoo Shin, and Jung~Hee Suh.
\newblock An optimal 9-point, finite-difference, frequency-space, 2-d scalar
  wave extrapolator.
\newblock {\em Geophysics}, 61(2):529--537, 1996.

\bibitem{hawkins2004problem}
Douglas~M Hawkins.
\newblock The problem of overfitting.
\newblock {\em Journal of chemical information and computer sciences},
  44(1):1--12, 2004.

\bibitem{mhaskar2016learning}
Hrushikesh Mhaskar, Qianli Liao, and Tomaso Poggio.
\newblock Learning functions: when is deep better than shallow.
\newblock {\em arXiv preprint arXiv:1603.00988}, 2016.

\bibitem{gao2019optimized}
Zhaoqi Gao, Zhibin Pan, Chen Zuo, Jinghuai Gao, and Zongben Xu.
\newblock An optimized deep network representation of multimutation
  differential evolution and its application in seismic inversion.
\newblock {\em IEEE Transactions on Geoscience and Remote Sensing},
  57(7):4720--4734, 2019.

\bibitem{jagtap2020conservative}
Ameya~D Jagtap, Ehsan Kharazmi, and George~Em Karniadakis.
\newblock Conservative physics-informed neural networks on discrete domains for
  conservation laws: Applications to forward and inverse problems.
\newblock {\em Computer Methods in Applied Mechanics and Engineering},
  365:113028, 2020.

\bibitem{wu2018efficient}
Zedong Wu and Tariq Alkhalifah.
\newblock An efficient helmholtz solver for acoustic transversely isotropic
  media.
\newblock {\em Geophysics}, 83(2):C75--C83, 2018.

\bibitem{yang2020multiparameter}
Qingjie Yang, Alison Malcolm, Bing Zhou, and Herurisa Rusmanugroho.
\newblock Multiparameter full-waveform inversion in fluid-saturated porous
  media.
\newblock In {\em SEG Technical Program Expanded Abstracts 2020}, pages
  900--904. Society of Exploration Geophysicists, 2020.

\bibitem{yang2020new}
Qingjie Yang, Bing Zhou, Mohamed~Kamel Riahi, and Mohammad Al-Khaleel.
\newblock A new generalized stiffness reduction method for 2-d/2.5-d
  frequency-domain seismic wave modeling in viscoelastic anisotropic media.
\newblock {\em Geophysics}, 85(6):1--70, 2020.

\bibitem{sitzmann2020implicit}
Vincent Sitzmann, Julien Martel, Alexander Bergman, David Lindell, and Gordon
  Wetzstein.
\newblock Implicit neural representations with periodic activation functions.
\newblock {\em Advances in Neural Information Processing Systems}, 33, 2020.

\end{thebibliography}

\end{document}